\documentclass[conference]{IEEEtran}
\usepackage{times}
\usepackage{cite}
\usepackage{algorithmic}
\usepackage[pdftex]{graphicx}
\usepackage{amsmath}
\usepackage{amsfonts}
\usepackage{amsthm,amssymb}
\usepackage{fixltx2e}
\usepackage{url}
\usepackage{subfig}
\usepackage{multirow}
\usepackage{rotating}
\usepackage{flushend}

\newtheorem{definition}{Definition}

\begin{document}
%
\title{Constraint Solvers \\ for User Interface Layout}
\author{%
{Noreen Jamil}%
\vspace{1.6mm}\\
\fontsize{10}{10}\selectfont\itshape
Department of Computer Science\\
University of Auckland\\
Private Bag 92019, Auckland, New Zealand\\
\fontsize{9}{9}\selectfont\ttfamily\upshape
\{njam031\}@aucklanduni.ac.nz\\
}
\maketitle
\begin{abstract}
Constraints have played an important role in the construction of GUIs , where they are mainly used to define the layout of the widgets.
Resizing behavior is very important in GUIs because areas have domain specific parameters such as form the resizing of windows.
If linear objective function is used and window is resized then error is not distributed equally. To distribute the error equally, a quadratic objective function is introduced.
Different algorithms are widely used for solving linear constraints and quadratic problems in a variety of different scientific areas.
The linear relxation, Kaczmarz, direct and linear programming methods are common methods for solving linear constraints for GUI layout.
The interior point and active set methods are most commonly used techniques to solve quadratic programming problems.
Current constraint solvers designed for GUI layout do not use interior point methods for solving a quadratic objective function subject to linear equality and inequality constraints.
In this paper, performance aspects and the convergence speed of interior point and active set methods are compared along with one most commonly used linear programming method when they are implemented for graphical user interface layout.
The performance and convergence of the proposed algorithms are evaluated empirically using randomly generated UI layout specifications of various sizes.
The results show that the interior point algorithms perform significantly better than the Simplex method and QOCA-solver, which uses the active set method implementation for solving quadratic optimization.
\end{abstract}
\vspace{1\baselineskip}
\section{Introduction}

Constraints are a suitable mechanism for specifying the relationships among objects.
They are used in the area of logic programming, artificial intelligence and UI specification.
They can be used to describe problems that are difficult to solve, conveniently decoupling the description of the problems from their solution.
Due to this property, constraints are a common way of specifying UI layouts, where the objects are widgets and the relationships between them are spatial relationships such as alignment and proportions.
In addition to the relationships to other widgets, each widget has its own set of constraints describing properties such as minimum, maximum and preferred size.

\begin{figure}[htb]
\centering
\includegraphics[width=1.1\columnwidth]{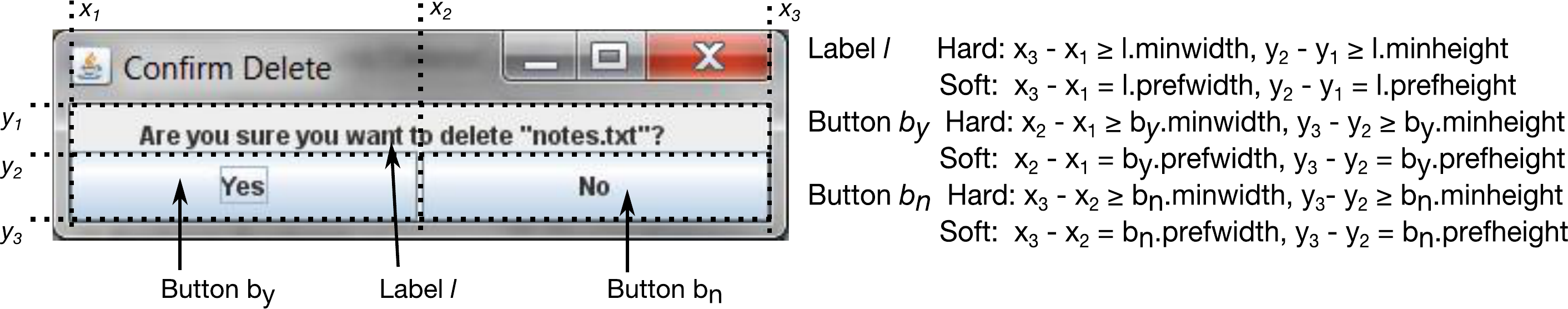}
\caption{Example constraint-based UI layout with hard and soft constraints} \label{fig:Gui-diagram}
\end{figure}

UI layouts are often specified with linear constraints~\cite{Weber:High-Level-Constraints}.
The positions and sizes of the widgets in a layout translate to variables.
Constraints about alignment and proportions translate to linear equations, and constraints about minimum and maximum sizes translate to linear inequalities.
Furthermore, the resulting systems of linear constraints are sparse.
There are constraints for each widget that relate each of its four boundaries to another part of the layout, or specify boundary values for the widget's size, as shown in Figure~\ref{fig:Gui-diagram}.
As a result, the direct interaction between constraints is limited by the topology of a layout, resulting in sparsity.

The Auckland Layout Model (ALM)~\cite{Lutteroth:Modular-Specification} enables the description
of Graphical User Interfaces (GUIs) in a constraint-based
manner.
Instead of placing widgets with absolute or relative
coordinates on a window the relations between them are
specified with constraints a GUI has to fulfill~\cite{Lutteroth:ordinal}.
Therefore it is easier to realize highly adaptable GUIs and achieve a
better modularity of GUI elements compared with common GUI
techniques such as Java's Gridbag Layout.

Constraints are the way to
formulate requirements on a specific GUI in ALM. A complete
layout in ALM is therefore defined by a set of areas (defined
by a set of tab stops) and a set of constraints.
With the given set of constraints an ALM layout manager has
to solve basically a system of linear equalities and inequalities.
However it is often the case in layout definitions that the
system is over-specified.
To cope with that problem ALM
introduces the notion of soft constraints~\cite{Weber:High-Level-Constraints}.

In contrast to the usual \emph{hard} constraints, which cannot be violated, soft constraints may be violated as much as necessary if no other solution can be found.
To solve
layouts which are defined with soft constraints and inequalities
it is not sufficient to solve a system of linear equations
but it is required to introduce a sort of optimization, namely
the minimization of the constraint-violation. The violation is
modeled with an additionally introduced penalty parameter for
each soft-constraint.

\begin{figure}[t]
  \centering
  \subfloat[Three buttons with quadratic objective function]
  {\includegraphics[width=1.2in]{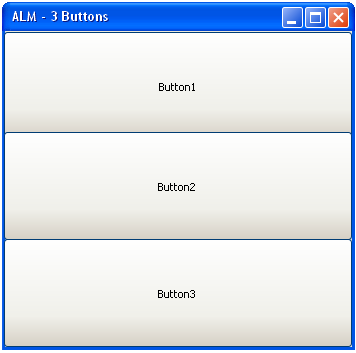}}%
   \qquad
  \subfloat[Three buttons with linear objective function]
  {\includegraphics[width=0.9in]{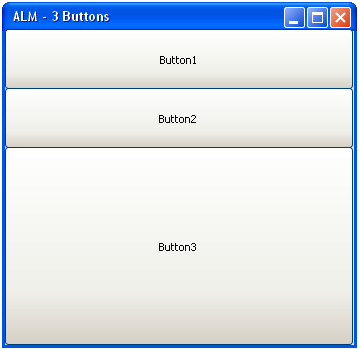}}
  \caption{\label{figure:twobuttons}Two different solving strategies for a simple two-button layout.}
\end{figure}

Current implementations of ALM use the
simplex algorithm for that task. However, with more complex GUI specifications the responsiveness
of the GUI decreases due to an increasing
computational effort. To increase the computational speed
a linear relaxation algorithm~\cite{Jamil:Linear-Relaxation} is currently used with a linear objective function.
One of the drawbacks of using linear objective function is the violation of soft constraints in a non-uniform way as shown in Figure~\ref{figure:twobuttons}, where only few constraints are violated but it is not precise which constraints are violated. This leads to the development of a quadratic objective function which minimizes the square of the deviations from a solution point to the defined constraints of a system.

Several different constraint-based GUI layout technologies have been developed and each of these technologies has their own peculiarities and requires specific knowledge.
By using these programmers and even end users can easily solve their problems with constraints since they have only to describe the problems.
The constraint idea was originated by Sutherland in 1960 who introduced Sketchpad~\cite{Sutherland:Sketchpad}, the first interactive graphical interface that solved geometric constraints.
Since then many constraint solvers have been developed and studied by the research community~\cite{Hosobe2000,Stuckey:arithmetic-constraints,Weber:High-Level-Constraints} and interest has increased with the recently introduced constraint-based layout model in the Cocoa API of Apple's Mac~OS~X\footnote{Cocoa Auto Layout Guide, 2012 \url{http://developer.apple.com}}.
Several different constraint-based GUI layout technologies have been developed and each of these technologies has their own peculiarities and requires specific knowledge.
By using these programmers and even end users can easily solve their problems with constraints since they have only to describe the problems.
The constraint idea was originated by Sutherland in 1960 who introduced Sketchpad~\cite{Sutherland:Sketchpad}, the first interactive graphical interface that solved geometric constraints.

Most researchers have
concentrated on developing and improving the performance of general
algorithms for the solution of many complex problems.
This is due to the rapid increase in the advancement in computer hardware (high speed processors, large memory etc).
While Cassowary~\cite{Badros:cassowary} was one of the pioneers in
developing algorithms for fast solution of GUI layout problems.
One of the challenges of the last few decades has been the construction of fast numerical
solution algorithms for solving GUI layout problems.

Recent developments in iterative methods have improved the efficiency of these methods.
The use of iterative methods has become ubiquitous in
recent years for solving sparse, real-world optimization
problems where direct algorithms are not suitable due to fill-in effects.
Unlike direct algorithms, which try to solve the problems finitely,
iterative methods start with a complete but preliminary task that is
not necessarily consistent. They improve this task in several
iterative steps until specified stopping criteria are satisfied. We can
get good approximate solutions by iterating the process, which is useful for practical applications especially if an efficient solution is required.

Much research has been carried out on constraint solving
techniques for linear programming problems. However, the linear programming technique tends to use an iterative method along with one step of a direct method.
Therefore, it is worth studying the potential of iterative algorithms because of their efficiency and capability to solve sparse linear constraint problems.

In this paper, we compare constraint solving techniques which are using quadratic and linear objective function.
These techniques were experimentally evaluated with regard to convergence and performance, using randomly generated UI layout specifications.
The results show that the interior point method is more efficient than the active set and simplex methods.
The simplex and active set methods have previously been used for UI layout.
In section~\ref{Conclusion}, we discuss related work.
A detailed description of quadratic programming and how systems of layout constraints can be solved using a quadratic objective functions is given in Section~\ref{Programming}.
Linear programming description and some overview of simplex algorithm is described in detail in Section~\ref{LinearProgramming}.
The methodology as well as the results of the evaluation can be found in Section~\ref{Evaluation}.
Section~\ref{Conclusion} finishes with conclusions and an outlook on future work.

\section{Related Work}\label{relatedwork}
Most of the research related to GUI layout deals with the various algorithms for solving constraint hierarchies. Research related to constraint based UI layout has provided results in the form of tools~\cite{Hosobe:Simplex-Based, Hosobe:scalable-linear-constraint} and algorithms ~\cite{Badros:cassowary,Stuckey:arithmetic-constraints} for specific tasks.
The latest work~\cite{Clemens:Aesthetics} on constraint based GUIs uses a
quadratic solving strategy which they find better than linear solving strategies. They~\cite{Clemens:Aesthetics} implemented
the active set method for solving a quadratic objective function subject to some linear constraints.
Baraf~\cite{Baraff:rigid-bodies} presents a quadratic optimization algorithm for
solving linear constraints in modelling physical systems.
QOCA~\cite{Stuckey:arithmetic-constraints} uses the active set algorithm for solving quadratic programming problem
for graphical user interface layout.

All constraint solvers for UI layout have to support over-constrained systems.
There are two approaches: weighted constraints and constraint hierarchies.
Weighted constraints are typically used with direct methods, while constraint hierarchies are used with linear programming.
Examples of direct methods for soft constraints are HiRise and HiRise2~\cite{Hosobe:Simplex-Based}.
Many UI layout solvers are based on linear programming and support soft constraints using slack variables in the objective function~\cite{Badros:cassowary,Stuckey:arithmetic-constraints,Weber:High-Level-Constraints}.

Many different local propagation algorithms have been proposed for solving constraint hierarchies in UI layout.
The DeltaBlue~\cite{Freeman:Delta-Blue}, SkyBlue~\cite{Skyblue:local-propagation-constraint-solver} and Detail~\cite{Hosobe:constraint-simultaneous} algorithms are examples in this category.
The DeltaBlue and SkyBlue algorithms cannot handle simultaneous constraints that depend on each other.
However, the Detail algorithm can solve constraints simultaneously based on local propagation.
All of the methods to handle soft constraints utilized in these solvers are designed to work with direct methods, so they inherit the problems direct methods usually have with sparse matrices.

None of the above discussed algorithms apply interior point methods for UI layout.

\section{Quadratic Programming} \label{Programming}

Quadratic programming deals with the optimization (minimization or maximization) of quadratic objective function that satisfies set of linear constraints.
\begin{definition}
Quadratic Programming is a problem which can be formulated as:
$$q(\overline{x}) := \frac{1}{2} \overline{x}^TQ\overline{x} - \overline{g}^T\overline{x} \rightarrow min$$
Subject to   \begin{math} Ax=b\end{math} and \begin{math}Cx\leq b\end{math} , \emph{$x\geq0$},\\
where the Hessian matrix $Q$ (positively semi-definite) of the quadratic function is an $n \times n$ quadratic matrix and \begin{math} x^{T} \end{math} is a vector transpose of x.
\end{definition}
\begin{math} Ax=b\end{math} and \begin{math} Cx\leq b\end{math} is a set of linear equality and inequality constraints where \emph{$x\geq0$} requires non-negativity conditions.
Generally, two approaches are used to solve quadratic programming problems.
These approaches are: interior point and active set.
The active set method is preferable if the QP problem is medium but the matrices are dense.
As large and sparse problems occur in user interface (UI) layout the method of choice is an interior point which is described in detail in the following section.
\subsection{Interior Point Method}
We use an interior point algorithm, the barrier method~\cite{Caroll:Barrier} for constraint based Graphical User Interfaces, which realizes
the error distribution with a quadratic objective function. It is an iterative method for solving constrained
optimization problems with inequality constraints. It is reasonably efficient and scalable up to thousand of constraints.
An interior point algorithm is one that starts inside the feasible area
and reaches the optimal solution through the interior of the
feasible region.

The key idea behind the algorithm is to start with a feasible point and a relatively large value of the
parameter r (where r is the barrier parameter).
A ``barrier'' function is used in order to define the feasible region of the
domain, i.e. to satisfy inequality constraints. As we proceed with the
iteration, the barrier function becomes steeper, so it forces to be
into the feasible region. Note that the feasible region is in both case an inner
region (that's where the word ``interior'' comes from), that is the inequality
constraints are always strict constraints $(<, not <=)$.
The barrier method uses an outer iteration in which the
barrier gets raised, and an inner iteration that solves this $i-th$ particular
problem, and so on until the final convergence.
The barrier term is added
to the objective function for a maximization problem and subtracted for a
minimization problem.
The details of algorithm are as follows.
\begin{itemize}
\item Step 1:
Initialization Step:
First choose an initial barrier parameter  $\mu> 1$, then a stopping
parameter $\epsilon > 0 $, and a strictly feasible $x$ that violates at least one constraint and
formulate the augmented objective function.
\item Step 2:
Repeat:
\item Step 3:
Centering Step:
Computing $x_{o}(t)$ by minimizing $tf_{o}+\phi$ subject to $Ax=b$, starting at $x$.
\item Step 4:
Updating Step:\\
Update. $x:=x_{o}(t)$.
\item Step 4:
Stopping Rule:\\
Quit if $m/t<\epsilon$
Increase $t$.  $t := \mu t$.
\end{itemize}
Here $f_{o}$ is the objective function, $\phi$ is the barrier function for the given
problem and $t$ is the weight of the barrier function in the $i-th$ sub problem.
As $t$ increases, the barrier function becomes steeper and better approximates
the inequality constraints. The starting point must be strictly feasible for all
of the constraints.

At each iteration the algorithm computes the central point $x_{o}(t)$ from the previously computed central point, and then increases $t$ by a factor of $\mu> 1$.
At the end we terminate the algorithm if $m/t<\epsilon$, otherwise we repeat the process.

Our solver uses the Gurobi library~\cite{Gurobi:Solver} for implementation of the interior point algorithm to solve our convex problem. It requires Gurobi to be installed and the Gurobi jar library in the build path.

\subsection{Active Set Method}
The active set method~\cite{Fletcher:1987:Active} is an iterative method for solving quadratic programming problems.
The idea behind the active set method is to solve a sequence of quadratic programming problem.
Each problem consists of a set of an objective function subject to equality constraints, which is known as the active set.
An active set contains equality constraints as well as inequality constraints (which are not active but must be fulfilled as equalities).
The active set method solves the quadratic programming problem by identifying the active set of its solution.
This method solves the hard constraints of the form:
\begin{align*}
A_i\cdot\overline{x} = \overline{b}_i \;\;\;\;\;\;\;\; & i \in equalities
\\
A_i\cdot\overline{x} \geq \overline{b}_i \;\;\;\;\;\;\;\; & i \in inequalities
\end{align*}
while minimizing the objective function:
$$q(\overline{x}) := \frac{1}{2} \overline{x}^TQ\overline{x} + \overline{g}^T\overline{x},$$
where $Q$ is symmetric.
The equality problem
$$A\cdot\overline{x} = \overline{b}$$

$$q(\overline{x}) := \frac{1}{2} \overline{x}^TQ\overline{x} + \overline{g}^T\overline{x} \rightarrow min$$

is analytically solvable by setting
$$\nabla q(\overline{x}) = 0$$
and solving the linear problem.
Steps for the active set algorithm are as follows:
\begin{itemize}
\item Step 1: Find a base solution for the hard constraints(linear system of inequalities)
$$A\cdot\overline{x} \geq \overline{b}$$

\item Step 2: Create an initial active set $A$ holding all soft constraints which satisfy the base solution as {\bf equality} constraints.

\item Step 3: From the base solution $\overline{x}$ find a new $\overline{x}_{new} = \overline{x} + \overline{\delta}$ which optimizes a objective function of the active set $A$ (soft constraints).
\end{itemize}

This implies for the hard constraints:
$$A\overline{x}_{new} = \overline{b}$$

$$\Leftrightarrow A\overline{x} + A\overline{\delta}= \overline{b}$$

$$\Rightarrow A\overline{\delta} = 0$$

{Get $\delta$ ($\overline{x}_{new} = \overline{x} + \overline{\delta}$)}

$\delta$ should lead us nearer to the optimal solution in respect to

$$q(\overline{x}) := \frac{1}{2} \overline{x}^TQ\overline{x} - \overline{g}^T\overline{x} \rightarrow min$$

$\Rightarrow$ let $\delta$ point to the searched minimum, means $\delta \rightarrow \nabla q(x^k)$

{Get $\delta$ ($\overline{x}_{new} = \overline{x} + \overline{\delta}$)}

This leads to the new quadratic sub problem:

$$k(\overline{x}) := \frac{1}{2} \overline{\delta}^TQ\overline{\delta} - \nabla q(x^k)^T\overline{\delta} \rightarrow min$$

subject to
$$A\overline{\delta} = 0$$

This means $\delta$ points as closely as possible in the direction of the derivation of $q$.

(this is solvable because it is an equality problem)

There are two cases in solving the problem.\\
{Case 1: $\delta = 0$ (remove a constraint)}
If $\delta = 0$ then remove a constraint that prevents further optimizing of solution:
This is the constraint with negative

$$min\lambda_i = \nabla{q(x^k)}_i$$

If there is no such a constraint (all $\lambda_i >= 0$) then stop.\\

{Case 2: $\delta \neq 0$ (add a constraint)}

calculate alpha such that:
$$\alpha^k = min(1, min\frac{b_i - (A\overline{x}^k)_i}{(A\overline{\delta}^k)_i})$$

If $\alpha < 1$ it means that the algorithm has not yet reached the constraint edge $b_i$
and if $\alpha = 1$ it means that the algorithm hits the constraint edge

\bigskip

If $\alpha < 1$ then add a constraint with smallest $\alpha$ (the constraint that the algorithm hits first)

Terminate algorithm if $\overline{x}^{k+1} = \overline{x}^k + \alpha^k \cdot \overline{\delta}$, otherwise repeat the process.

\subsubsection{Advantages of Interior Point and Active Set Methods}
The interior point method has certain advantages over the active set method.
The interior point method is simple to implement and efficient.
It is efficient especially for sparse matrices, i.e.\ matrices where the number of non-zero elements is a small fraction of the total number of elements in the matrix~\cite{Stephen:Convex}.

For general nonlinear optimization problems, barrier methods are among the most powerful classes of
algorithms~\cite{Nick:Numerical}.
The statement that supports this fact is that these methods
will converge to at least a local minimum in most cases, even if the constraints and objective functions do not have convexity characteristics. They
work well even in the presence of spinode and similar form that can
mystify other approaches.

An interior point is less sensitive to problem size whereas active set adds  a combinatorial element to the identification of the
active set and the solution of the quadratic programming, and as a result computational effort can increase with the problem
size~\cite{Dulce:Barrier}.

Considering these advantages, we choose the interior point method for solving GUI layout problems.
\section{Linear Programming} \label{LinearProgramming}
Linear programming~\cite{Hanif:Lp-NF} deals with the optimization (minimization or maximization) of an objective function that satisfies a set of constraints.

A specification as a linear program is trivially in general more
expressive than a specification as a system of linear equations and
inequalities. The specification as a system of linear equations and inequalities is a special case of linear programming
with the trivial
objective function $0$.

\begin{definition}
Linear Programming is a problem which can be formulated in standard form as:

Minimize  \begin{math} c^{T}x \end{math}

Subject to   \begin{math} Ax=b\end{math} ,\emph{$x\geq0$},\\
where  \begin{math} c^{T}x \end{math} is a linear objective function.
\end{definition}

\begin{math} Ax=b\end{math} is a set of linear constraints and \emph{$x\geq0$} requires non-negativity conditions.

In the maximization case, minimizing \begin{math}c^{T}x\end{math} is equivalent to maximizing \begin{math}-c^{T}x\end{math}.
Inequality constraints are included because \begin{math} A^{'}x\leq b\end{math} or \begin{math} A^{''}x \geq b\end{math} is equivalent to \begin{math}Ax=b\end{math} by including slack and surplus variables as required.

Linear Programming is mostly used in constrained optimization. A
large number of optimization problems are LPs having hundred of
thousands of variables and thousands of constraints. With the
recent advancement in computer technology these problems can be
solved in practical amounts of time. The most common algorithm to solve linear programming problems is called the simplex method, which is described below.

\subsection{Simplex Method} \label{SimplexMethod}
The simplex method~\cite{Dantzig:Linear-Programming} also known as the simplex technique or simplex
algorithm was developed in 1947 by the American mathematician George
B.Dantzig. It is an iterative method and makes use of Gauss-Jordan elimination techniques. It has the advantage of being universal, i.e. any linear
model for which a solution exists can be solved by it. It is defined as
an algebraic process for solving linear programming problems.

The simplex method is an iterative process that starts at a feasible
corner point(normally the origin) and systematically moves from one
feasible extreme point to another, until an optimal point is
eventually reached.\\
The simplex method usually has two stages, called phase-I and phase-II.\\
In phase-I, the algorithm finds a basic feasible solution.\\
In phase-II, the algorithm searches for an optimal solution.
In phase-I, slack variables(a slack variable is added to a constraint to turn an inequality
into an equation) are introduced to find a value of the decision variables
where all the constraints are satisfied. Once a basic feasible solution is found, the search for an optimal solution can start.
In phase-II, the algorithm moves from one extreme point to another to find the optimal solution.
The next extreme point will be chosen such that the search direction is in the steepest feasible direction.
This process continues until the optimum solution is reached.

\section{Experimental Evaluation} \label{Evaluation}
In this section we present an experimental evaluation of the proposed algorithms.
We conducted two different experiments to evaluate (i) the convergence behavior, (ii) the performance in terms of computation time.
The experiments were conducted as follows.

\subsection{Methodology}
For both experiments we used the same computer and test data generator, but instrumentalized the algorithms differently.
We used the following setup:
a desktop computer with Intel Core 2 Duo 3GHz processor under Windows 7, running an Oracle Java virtual machine.
Layout specifications were randomly generated using the test data generator described in~\cite{Weber:High-Level-Constraints}.
For each experiment the same set of test data was used.
The specification size was varied from 4 to 2402 constraints in increments of 4 constraints (2 new constraints for positioning and 2 new constraint for the preferred size of a new widget).
For each size 10 different layouts were generated resulting in a total of 6000 different layout specifications which were evaluated.

This test data served as input for our algorithms.
We conducted two experiments.
In the first experiment we investigated the convergence behavior of the algorithms.
We measured for each algorithm the number of sub-optimal solutions.
A solution is sub-optimal if the error of a constraint (the difference between right hand- and left hand sides) is not smaller than a given tolerance.

In the second experiment we measured the performance in terms of computational time (\(T\)) in milliseconds (ms), depending on the problem size measured in the number of constraints (\(c\)).
Each of the proposed algorithms was used to solve each of the problems of the test data set and the time was taken.
As a reference, all the generated specifications were also solved with, linear and quadratic constraint solving methods, QOCA solver, the interior point and Simplex methods.
For comparison purposes we selected QOCA solver that has been implemented for solving convex quadratic programming problem for UI layout.

The QOCA solver uses largely
distinct weights (e.g., 1, 1000, and 1000000) to handle constraint
hierarchies.
Whereas, we handle soft constraints in our problem formulation by introducing a slack variable
per constraint. In the objective function the values of these slack variables are squared and weighted by the penalties of the corresponding
constraint.
\subsection{Results}
In the first experiments we investigated the convergence behavior of all algorithms.
We found that both algorithms converge in the end.
This result is obvious since the algorithms are designed to find a solvable subproblem.

\begin{figure*}[htb]
\centering
\includegraphics[width=4.1 in, height=3.1in]{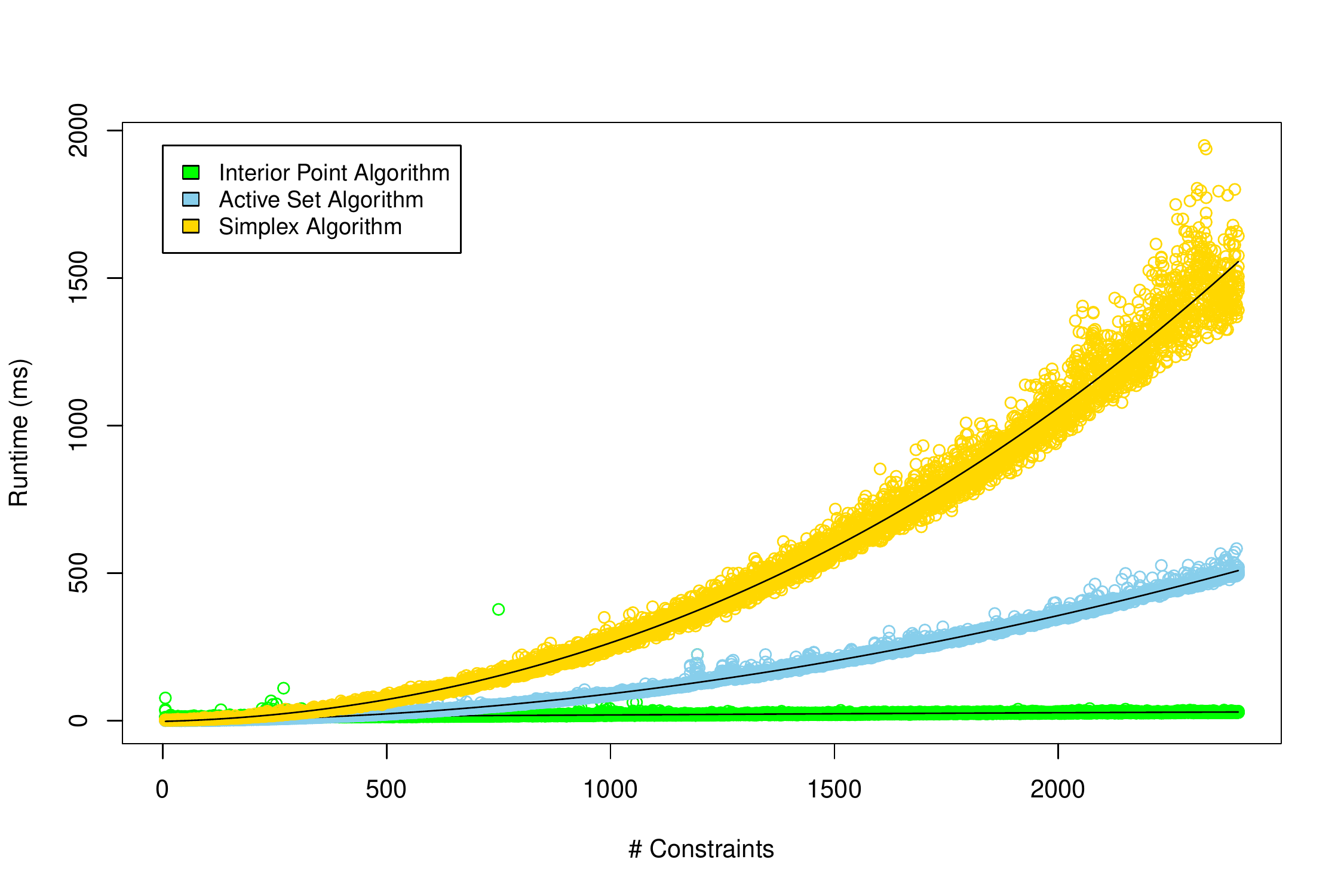}
\caption{Performance comparison of Interior Point, Active Set and Simplex Algorithms} \label{fig:random}
\end{figure*}

In the second experiment we investigated the computational time behavior of all algorithms.
To figure out the trend of the performance of the algorithms we defined some regression models (linear, quadratic, log, cubic).
We found that the best fitting model is the polynomial model
\[
T = \beta_0 + \beta_1c + \beta_2c^2 + \beta_3c^3 + \epsilon
\]
which gave us a good fit for the performance data.
Key parameters of the models are shown in Table~\ref{tab:reg1}; a graphical representation of the models can be found in Figures~\ref{fig:random}. Table~\ref{tab:symbs} explains the symbols used.
\begin{table}[ht]
\begin{center}
\begin{tabular}{rl}
  \hline
 Symbol & Explanation \\
  \hline
 \(\beta_0\) & Intercept of the regression model\\
 \(\beta_{1-3}\) & Estimated model parameters\\
 \(c\) & Number of constraints\\
  \(T\) & Measured time in milliseconds\\
   \(R^2\) & Coefficient of determination of the estimated regression models\\
\hline
\end{tabular}
\caption{Symbol Table}
\label{tab:symbs}
\end{center}
\end{table}

\newcommand{\sign}[1]{\textsuperscript{#1}}
\begin{table*}[t]
\begin{center}
\begin{tabular}{rlllll}
  \hline
\multicolumn{1}{c}{\bf Strategy}                       &\multicolumn{1}{c}{\(\beta_0\)} &\multicolumn{1}{c}{\(\beta_1\)} & \multicolumn{1}{c}{\(\beta_2\)}  & \multicolumn{1}{c}{\(\beta_3\)} & \(R^2\)  \\
  \hline
Interior Point Algorithm & \(\;\;\, 1.179e^{+01}\)\sign{***}& \( 6.645e^{-03}\)\sign{***} & \(\;\;\,  1.443e{-06}\)\sign{***} & \(\;\;\, -4.868e^{-10}\)\sign{***}& \(0.4145\)\\
Active Set Algorithm &\(\;\;\, 1.225e^{+00}\)\sign{***} & \( -3.273e^{-03}\)\sign{***} & \(\;\;\,  9.595e^{-05}\)\sign{***} & \( -2.765e^{-09}\)\sign{***}& \(0.9971\)\\
Simplex Algorithm &\( -2.491\)\sign{***} & \(\;\;\, 3.924 \cdot 10^{-02}\)\sign{***} & \(\;\;\, 2.079 \cdot 10^{-04}\)\sign{***} & \(\;\;\, 1.904 \cdot 10^{-08}\)\sign{***} &\(0.9900\) \\
\hline
\multicolumn{6}{l}{Significance codes: \sign{***} \(p<0.001\)} \\
\end{tabular}
\caption{Regression models for the different solving strategies}
\label{tab:reg1}
\end{center}
\end{table*}

For some strategies some parameters do not have a significant effect.
That can be interpreted as the complexity of the algorithm not following a certain polynomial trend.
As the graphs indicate, interior point exhibits better performance than active set simplex algorithms.

Figure~\ref{fig:random} compares interior point, active set and simplex algorithms.
Generally, the active set method is slower than the interior point method for solving convex quadratic programming problem for UI layout because of changes in an active set estimate combinatorially. The reason behind slowness of simplex algorithm is one Gauss-Jordan elimination step per iteration, i.e.\ using a direct method.

\subsection{Discussion}
The performance results show that the interior point method is faster than the active set method.
One reason why the active set method is slow is that it
adds a combinatorial element in identifying the
active set and the solution of the quadratic programming, and this can increase computational effort for solving the problem.
Whereas the interior point method is independent to the problem size.

Even though active set methods have the advantage of warm starts but this approach can be inefficient as explained below.
Most of the complex operations (adding and deleting a constraint to the active set, estimates for Lagrange multiplier, computing search directions etc) involved in the make up of algorithm.
On the other hand the most complex operation in the interior point method is the solution of a linear system and this operation is fairly simpler than the active set method.

An active set takes large number of iterations to converge for large sparse problems whereas an interior point takes less.

A plausible reason the simplex method is slower is that even though it is an iterative method but it uses one direct method solving step per iteration.
As direct methods suffer from fill-in effects when solving sparse systems, which is generally a disadvantage compared to iterative methods in this case.
\section{Conclusion} \label{Conclusion}

In this paper we have compared the performance of interior point and active set methods for solving
convex programming problems for constraint based GUI layouts. We have compared the speed and convergence of these methods.
We found that the interior point method is more efficient for large sparse problems than the active set method.
We also compared the performance of simplex algorithm and found that it is slowest than interior point and active set methods.
The work presented in this paper lays a foundation for the application of iterative methods for solvers of constraint-based UIs.
We identify the following future work in that area.

First, some applications in constraint-based UIs could benefit from the possibility to formulate non-linear constraints.
Integrating the solving of non-linear constraints into the framework of the Gauss-Seidel method would extend the application domain of our algorithms.

Second, there is room for improvement in the deterministic pivot assignment algorithm in linear relaxation.
The results of the experiment indicate that an optimal pivot assignment can have a huge effect on the speed of convergence.
Currently, deterministic pivot assignment only takes the influence of coefficients of constraints into account.
The inferior performance of this selector compared to a purely random one indicates that there are other factors that have an effect on convergence.
One such factor is the order of the constraints.

Third, we have proven the convergence theorem for the Gauss-Seidel method for the case of non-square, row-dominant coefficient matrices.
However, our experimental evaluation indicates that linear relaxation converges for some UI layout problems which do not fully satisfy the row-dominance criterion.
A weaker convergence criterion would be very insightful and could lay the basis for further improvement of the algorithms.

With the contributions mentioned above we have demonstrated that iterative method can efficiently be used for solvers for constraint-based UIs.
With the algorithms presented in this paper it is possible to bring the benefits of solving sparse matrices efficiently with iterative methods to the domain of UI layout.

\bibliographystyle{IEEEtran}
\bibliography{bibliography}
\end{document}